\documentclass[conference]{IEEEtran}
\IEEEoverridecommandlockouts

\usepackage{cite}
\usepackage{amsmath,amssymb,amsfonts}
\usepackage{algorithm}
\usepackage{algpseudocode}
\usepackage{graphicx}
\usepackage{textcomp}
\usepackage{xcolor}
\usepackage{booktabs}
\usepackage{multirow}
\usepackage{url}
\usepackage{hyperref}
\graphicspath{{figures/motivation/}{figures/exp/}{figures/architecture/}{figures/}}

\hypersetup{
    colorlinks=true,
    linkcolor=blue,
    citecolor=blue,
    urlcolor=blue
}

\newcommand{\hi}[1]{\vspace{.25em}\noindent\textbf{#1}}
\newtheorem{definition}{Definition}

\begin{document}

\title{RECAST: A Region-Scoped Adaptive Index for Exact Similarity Search}

\author{
\IEEEauthorblockN{
Yining Liu$^{*}$, Rui Mao$^{\dagger\,\boxtimes}$
}
\IEEEauthorblockA{
$^{*}$Beijing Institute of Technology, Zhuhai, China\\
$^{\dagger}$Shenzhen University, Shenzhen, China\\
Email: lynnn@bit.edu.cn, mao@szu.edu.cn\\
$^{\boxtimes}$Corresponding author
}
}

\maketitle

\begin{abstract}
Similarity search in metric spaces is routinely used in applications including bioinformatics, data mining, and recommender systems. Exact similarity search is dominated by distance computations, while real query streams often concentrate in specific regions rather than spreading uniformly across the space. Pre-built indexes are constructed before the query stream; if queries concentrate in regions poorly served by the fixed index structure, pruning remains weak and cannot improve online. Adaptive indexes such as AV-tree build the index from distances computed while answering queries, but discard most of the distances each query computes and do not effectively organize the retained distances for reuse, so valuable computation is not fully exploited to serve future queries in the same region.
These limitations raise three challenges: the query-concentrated region is not known in advance and may change over time; distances already computed while answering earlier queries must be accumulated and reused only where they remain effective; and outdated distances from one region should not add overhead to queries in another. We propose RECAST, a region-scoped adaptive index for exact similarity search. RECAST maintains query regions, accumulates distances already computed while answering earlier queries (paid distances) within each region for exact pruning, and uses changes in their pruning effect to infer whether incoming queries are still concentrated in the current region. When queries shift, RECAST recursively dispatches query work to child regions, so paid distances are accumulated and reused only where they remain effective. On five real-world datasets under four workload patterns, RECAST achieves consistently lower cumulative cost than the adaptive baseline and most pre-built baselines, reducing per-query distance computations by up to 64\% and query time by up to 46\% compared with the state-of-the-art adaptive baseline AV-tree.
\end{abstract}

\begin{IEEEkeywords}
similarity search, metric space, adaptive indexing
\end{IEEEkeywords}

%==========================================================
\section{Introduction}\label{sec:intro}
%==========================================================

% Paragraph 1: Problem + importance + applications
Similarity search in metric spaces is a fundamental operation in bioinformatics, multimedia retrieval, recommendation, and vector databases. Given a query object and a radius, an exact range query returns all database objects within that distance. Since objects in a metric space may lack coordinates and distance functions can be expensive, exact search is dominated by distance computations. In real workloads, queries often concentrate in specific regions rather than spreading uniformly across the space. For example, a researcher exploring a molecular database may first search intensively around one compound family and then shift to another. In each concentrated phase, nearby queries repeatedly encounter similar hard-to-prune candidates. This creates an opportunity to accelerate future queries by reusing what previous queries have already computed.

% Paragraph 2: Limitations (concise, three-fold)
\hi{Limitations of existing approaches.}
Existing indexes for exact similarity search fall into two groups: those that build a fixed structure before the query stream, and those that adapt query by query. Pre-built indexes such as LAESA~\cite{mico1994laesa}, GNAT~\cite{brin1995gnat}, SAT~\cite{navarro2002sat}, and the learned index LIMS~\cite{tian2022lims} optimize global separability without future query information. When queries concentrate in regions poorly served by the fixed structure, these methods cannot adapt online and pruning stays weak. AV-tree~\cite{lampropoulos2023avtree} is the closest adaptive baseline for exact similarity search in general metric spaces. It builds the index incrementally during queries~\cite{idreos2007cracking}, but each query contributes only a narrow refinement of the tree: it compresses all cracking distances into a single global tree and keeps at most one cached distance per object, so leaf-level pruning is bounded by a single pivot. Many distances paid while answering a costly query are not turned into a richer local pruning state, and later nearby queries may still revisit similar hard candidates. AV-tree also cannot distinguish whether a costly query produces reusable distances, indicates a saturated region, or signals a new task.

\begin{figure}[t]
\centering
\includegraphics[width=0.6\columnwidth]{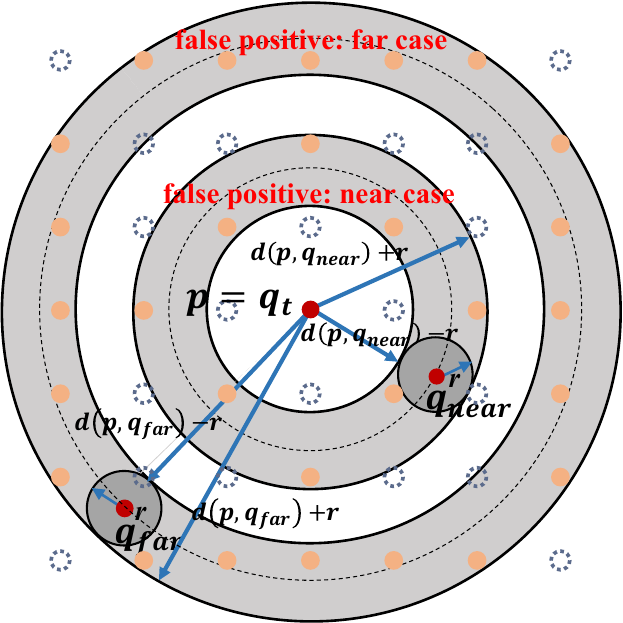}
\caption{Pivot pruning on uniformly distributed data.}
\label{fig:annulus}
\end{figure}

Figure~\ref{fig:annulus} illustrates this effect on uniformly distributed data. Given a pivot $p$ (a historical query) and a search radius $r$, the objects that cannot be pruned by the triangle inequality lie in an annulus centered at $p$ with inner radius $\max(0, d(q,p){-}r)$ and outer radius $d(q,p){+}r$. In the figure, gray dashed circles are pruned objects and orange dots are unprunable false positives that must be verified by exact distance computation. For a nearby query $q_{\text{near}}$, the annulus is narrow and contains few objects (strong pruning). For a distant query $q_{\text{far}}$, the annulus shifts outward and, in typical data distributions, covers more objects (weak pruning). This asymmetry is the source of adaptive benefit: an index that accumulates paid distances in a region where queries concentrate is often more effective than a pre-built global structure, because each stored distance is closely aligned with the query neighborhood it serves.

% Paragraph 3: Three challenges (problem-level, detailed)
\hi{Challenges.}
When queries concentrate in specific regions, the distances that a query is forced to compute expose which objects are hard to prune, and this difficulty often recurs in nearby future queries. Exploiting this regularity while answering queries, without seeing the future, raises three difficulties.

\hi{Challenge 1: The query-concentrated region is not known in advance and may change.}
The benefit of adaptive indexing comes from query concentration, but the index cannot observe where queries concentrate. It sees only the distances each query computes: how many objects were checked, how many were false positives, how many were pruned by the existing structure, and how these counts change over time. The same high cost can arise from different causes. The index may not yet have enough pruning power in this region, so continued learning would help. The available pruning mechanisms may already be near their limit in this region, so further effort yields little gain. The query may have moved to a new region where previously accumulated information does not apply. Or the query parameters may have changed. The challenge is to infer from these observable costs whether historical information is still relevant to the current query region.

\hi{Challenge 2: An online index must decide when to exploit accumulated information and when to adapt to a new region.}
An online adaptive index must decide how each new query should change the structure. If the index changes the structure on every query, for example by unconditionally cracking large leaves as AV-tree does, repeated reorganization disrupts the structure learned under a stable workload, and the small modification from a single query may not accumulate into a strong local pruning state. If the index is too conservative and keeps the old structure, the structure fails on new queries when the workload drifts or shifts to a new region. The challenge is to keep learning when queries remain concentrated, to stop expanding when a region is saturated, and to switch to a new scope when the query focus shifts. Distinguishing these regimes requires reading the cost signals that each query exposes.

\hi{Challenge 3: Historical information has scope and can pollute new queries.}
The distances paid by a past query are most effective for future queries in the same region. For queries in a different region, their pruning power is often weak and may vanish entirely: the stored distances fail to prune the new candidates, but the index still pays the cost of checking them. If a partition splits a region but leaves objects whose distance to the split center was not computed in the parent, different query tasks accumulate in the same scope and the observable cost signals become ambiguous. The challenge is to maintain the correct scope for historical information so that it stays where it is useful and the rest forms a new scope that can learn on its own.

\begin{figure}[t]
\centering
\includegraphics[width=\columnwidth]{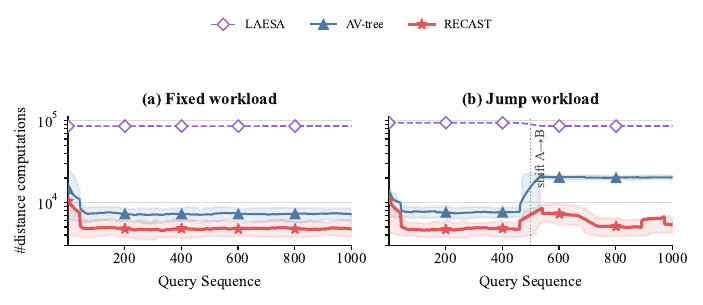}
\caption{Per-query distance computations on uniform20d under fixed and jump workloads.}
\label{fig:motivation_curves}
\end{figure}

Figure~\ref{fig:motivation_curves} confirms this on uniform data. Under a fixed workload (Figure~\ref{fig:motivation_curves}(a)), both adaptive methods reduce per-query cost over time as accumulated distances provide pruning for subsequent queries in the same region; LAESA's cost stays flat because its global pivots do not learn from the query stream. Under a jump workload (Figure~\ref{fig:motivation_curves}(b)), the cost of both adaptive methods spikes when the query focus shifts at query~500, because the distances accumulated in region~A are far from region~B and its pruning power degrades as the annulus analysis predicts. AV-tree's spike reaches $2.88\times$ its pre-shift cost and recovers slowly ($0.929\times$), because the tree structure shaped by queries in region~A provides little pruning for queries in region~B, and each leaf retains at most one cached distance, limiting the multi-pivot pruning that could accelerate re-learning. RECAST's spike is smaller ($1.75\times$) and recovery is faster ($0.742\times$), because its region-scoped design allows region~B to accumulate its own pivot distances independently of region~A.

% Paragraph 4: Our approach
\hi{Our approach.}
We observe that the distances a query is forced to compute serve a dual role. They are paid distances that can prune future candidates through the triangle inequality, and they are a signal of whether the current structure fits the current query region. RECAST builds on this observation.

RECAST is an adaptive index for exact similarity search that treats each query region as an independent recursive online task. Each region maintains its own paid pivot table for object-level pruning. The same distance-computation statistics serve as runtime signals of regional task difficulty: the index reads whether the count of checked objects is falling, flat, or rising to assess whether its historical information remains relevant (addressing Challenge~1). When these signals indicate that the current structure no longer fits, RECAST does not reorganize at once but runs a shadow test over subsequent queries. The test checks whether a candidate partition would reduce later distance computations relative to the current pruning path, and the partition is committed only if the test confirms a net saving, so regions that are saturated or still learning are naturally excluded from reorganization. The same cost signal also drives pivot retention within a region: when the cost is falling, the region conservatively keeps its accumulated distances; when the cost spikes, the region prefers evicting older pivots to make room for new distances. These two adaptive decisions share the same query-cost observations rather than relying on independent heuristics (addressing Challenge~2). After a partition, objects whose distance to the split center was not computed are collected into a residual child region with its own table. This prevents outdated distances from adding overhead to new queries and keeps these objects in a scope that can continue to learn and to partition (addressing Challenge~3).

% Paragraph 5: Experimental evidence
\hi{Experimental evidence.}
On five real-world datasets under four workload patterns, RECAST achieves consistently lower cumulative cost than the adaptive baseline and most pre-built baselines, reducing per-query distance computations by up to 64\% and query time by up to 46\% compared with the state-of-the-art adaptive baseline AV-tree (Section~\ref{sec:exp}).

% Paragraph 6: Contributions
\hi{Contributions.}
Our contributions can be summarized as follows.
\begin{itemize}
    \item \textbf{Recursive region-adaptive indexing framework.} We propose RECAST, a region-scoped adaptive index for exact similarity search in metric spaces. RECAST models each query region as an independent recursive online task. Each region maintains its own paid pivot table for object-level pruning and reuses query-paid distances as signals of regional task difficulty. Objects not covered by a partition form independent child regions that continue to learn, so the entire dataset remains continuously adaptive (Section~\ref{sec:method:overview}).
    \item \textbf{Signal-driven adaptive pivot management} (addresses Challenge~2). We design a per-region paid pivot table whose retention and eviction are driven by the region's cost signal rather than by a fixed lifecycle rule. The same signal that distinguishes ``still learning'', ``saturated'', and ``shifted'' regimes governs both pivot retention and partition timing, so pivot management and partition decisions share a unified adaptive framework (Section~\ref{sec:method:region}).
    \item \textbf{Cost-signal-driven partition algorithm with shadow validation.} We introduce a partition decision algorithm that reads the trend of distance computations in a region, including the checked count, the false-positive ratio, the pruning rate, and their evolution, to determine when the current structure no longer fits the query workload. A candidate partition is committed only after a shadow test over subsequent queries confirms that it reduces distance computations beyond the current pruning path (Section~\ref{sec:method:dispatch}).
    \item \textbf{Extensive experiments.} On five datasets under four workload patterns (20 combinations), RECAST reduces per-query distance computations by up to 64\% and query time by up to 46\% compared with AV-tree, winning all 20 combinations in distance computations. Four ablation studies confirm the individual contributions of each component (Section~\ref{sec:exp}).
\end{itemize}

%==========================================================
\section{Problem Statement}\label{sec:problem}
%==========================================================

We define the metric space, the similarity query, and the adaptive setting we study. Table~\ref{tab:notation} lists the frequently used notation.

\hi{Metric space.}
A metric space is a pair $(M, d)$, where $M$ is a domain of objects and $d : M \times M \rightarrow \mathbb{R}_{\geq 0}$ is a distance function. For all $x, y, z \in M$, $d$ satisfies identity ($d(x,x)=0$), positivity ($x \neq y \Rightarrow d(x,y) > 0$), symmetry ($d(x,y)=d(y,x)$), and the triangle inequality ($d(x,z) \leq d(x,y) + d(y,z)$). Distance functions used in practice include Manhattan and Euclidean distance on numerical features, edit distance on strings, and quadratic-form distance on color histograms. Evaluating $d$ can take milliseconds, so the number of distance computations is the dominant cost of a query. We index a dataset $O \subset M$ with $|O| = n$ objects.

\begin{table}[t]
\caption{Frequently used notation}\label{tab:notation}
\centering
\small
\begin{tabular}{ll}
\toprule
\textbf{Symbol} & \textbf{Meaning} \\
\midrule
$(M, d)$ & metric space and distance function \\
$O$, $n$ & dataset and its cardinality $|O|$ \\
$o$, $x$ & a data object in $O$ \\
$q$, $r$ & a query object and its search radius \\
$R$ & a region, a subset of $O$ with its own table \\
$P_R$ & the paid pivot table of region $R$ \\
$p$ & a pivot (a past query center or split center) \\
$s$ & a split center of a region \\
$\theta$ & the leaf-size threshold \\
\bottomrule
\end{tabular}
\end{table}

\hi{Range query.}
Given a query object $q \in M$ and a search radius $r \geq 0$, a range query seeks all objects in $O$ within distance $r$ from $q$~\cite{lampropoulos2023avtree, chen2022survey}.

\begin{definition}[Range Query]\label{def:range}
Given a dataset $O$, a query object $q \in M$, and a radius $r \geq 0$, a range query returns
\[
\mathit{RQ}(q, r) = \{\, o \in O \mid d(q, o) \leq r \,\}.
\]
\end{definition}

Given a positive integer $k$, a $k$-nearest-neighbor ($k$NN) query seeks the $k$ objects in $O$ with the smallest distance to $q$. A $k$NN query can be reduced to a sequence of range queries with a shrinking radius. Our implementation includes a basic $k$NN variant; adapting the cost signal to the dynamic radius for stronger performance is left for future work. This paper focuses on range queries.

\hi{Adaptive setting.}
We study similarity search in an adaptive setting. The dataset $O$ is given, but no index over $O$ is built in advance. Queries $q_1, q_2, \ldots$ arrive in sequence, and the index must answer each $q_i$ before $q_{i+1}$ arrives. Distances computed while answering a query may be kept and reused for later queries, but the index never runs a separate construction phase that computes distances unrelated to a pending query. We measure the cost of a query by the number of distance computations it performs, and report query time as a secondary measure.

%==========================================================
\section{The RECAST Index}\label{sec:method}
%==========================================================

We derive the design of RECAST from three observations about the nature of distance computations in adaptive indexing for exact similarity search. These observations are properties of the problem setting, not of any particular index.

\hi{Observation 1: Every distance computation has dual value.}
When an adaptive index computes $d(x, p)$ to answer a query, this distance serves two purposes. It determines whether $x$ belongs to the result, and it becomes a reusable fact: for any future query $(q, r)$, the triangle inequality gives $d(q, x) \geq |d(x, p) - d(q, p)|$, so $x$ can be pruned without computing $d(q, x)$ if this lower bound exceeds $r$. The distance is both the cost of the current query and a reusable paid distance for future queries. This dual value exists as soon as the distance is computed.

\hi{Observation 2: The reuse value of a distance is local, not global.}
The pruning power of a stored distance $d(x, p)$ for a future query $(q, r)$ depends on $d(q, p)$. When $q$ is close to $p$, the objects that $p$ cannot prune form a narrow band, and most others are pruned. When $q$ is far from $p$, the band widens and the stored distance loses its pruning power. In high-dimensional spaces where distances concentrate, the band may cover most objects even for nearby queries, limiting all pivot-based pruning. Nevertheless, the dependence on $d(q,p)$ holds in all settings: distances paid by queries in one region are most useful for future queries in the same region. A single global table that mixes distances from different regions dilutes the locally useful pruning power. More precisely, given pivot $p$ and query $(q, r)$, the objects that cannot be pruned by $p$ are those with $d(x,p) \in [\max(0, d(q,p){-}r),\; d(q,p){+}r]$. As $d(q,p)$ grows, this interval shifts outward and typically contains more objects, so the fraction of candidates that survive the pivot increases (Figure~\ref{fig:annulus}). This is why distances computed by nearby queries provide strong pruning while distances from distant queries contribute little.

\hi{Observation 3: The number of distances a query still computes directly measures structure fitness.}
After an adaptive index has accumulated some structure, each new query computes fewer distances than a linear scan. The count of distances still computed, relative to the region size and its recent trend, measures how well the current structure serves the current workload. A falling count means the structure is learning. A flat or rising count means the structure may have reached its limit or the workload may have shifted. We call this count the \textit{cost signal}; it is a free byproduct of query execution.

\hi{Design implications.}
These observations lead to three design choices. From Observation~1, distances computed during a query that meet the admission threshold are retained as a trial pivot column in the region where the query was processed (Section~\ref{sec:method:region}). From Observation~2, we scope each pivot table to its own region, and when a region no longer fits the workload, we create new regions with their own tables (Section~\ref{sec:method:overview}). Observation~3 is used at two levels: the same cost signal drives both pivot retention in Section~\ref{sec:method:region} and partition timing in Section~\ref{sec:method:dispatch}, so the two adaptive decisions share a single mechanism rather than independent heuristics.

%----------------------------------------------------------
\subsection{Index Structure and Query Flow}\label{sec:method:overview}
%----------------------------------------------------------

\begin{figure*}[t]
\centering
\includegraphics[width=0.85\textwidth]{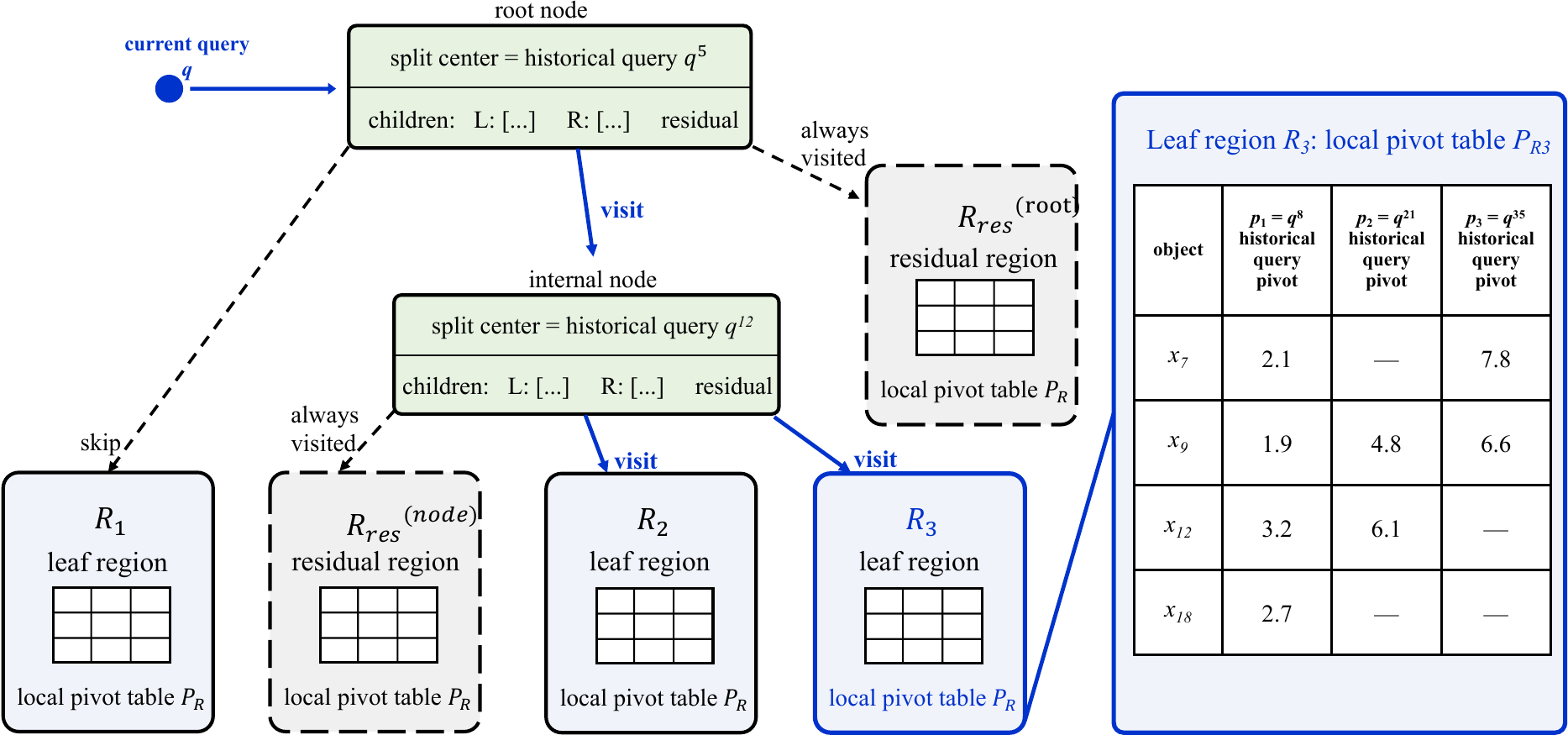}
\caption{RECAST query processing. Each leaf region maintains its own pivot table $P_R$. All split centers and pivot centers are historical queries.}
\label{fig:architecture}
\end{figure*}

This subsection describes how RECAST organizes the dataset and processes queries. Figure~\ref{fig:architecture} illustrates the query flow on a small example.

\hi{Region.}
RECAST organizes the dataset as a tree of regions. A region $R$ holds four components: an object scope $O_R \subseteq O$, a \textit{paid} pivot table $P_R$ (so called because every distance in $P_R$ was computed as part of answering a past query, not in a separate construction phase), query statistics $H_R$ that track recent distance-computation counts, and, if $R$ has been partitioned, three children (left, right, and residual). Every region runs the same logic. The root region covers the whole dataset. A leaf region covers a small subset. There is no separate global structure.

\hi{Running example.}
Figure~\ref{fig:architecture} shows the state of RECAST after processing several queries that triggered two partitions. Initially, all objects reside in a single root region with an empty pivot table. As queries arrive, the root accumulates pivot columns and its cost signal stabilizes. When query $q^5$ arrives, the candidate selection conditions (Section~\ref{sec:method:dispatch}) are met and a shadow test confirms that splitting around $q^5$ would reduce future distance computations. The root is partitioned: objects whose distance to $q^5$ falls below the median go to $R_1$, those above go to the right child, and objects whose distance was not computed go to the residual child $R_{\text{res}}^{(\text{root})}$. The split center $q^5$ is installed as the first pivot in both $R_1$ and the right child at no extra cost. The right child then accumulates its own pivots from subsequent queries. When query $q^{12}$ triggers a second partition of the right child, the same process produces $R_2$, $R_3$, and $R_{\text{res}}^{(\text{node})}$; existing pivots in the right child, such as those from $q^8$, are inherited by the new children.

When the current query $q$ arrives (left side of Figure~\ref{fig:architecture}), RECAST routes it through the tree. At the root, $d(q, q^5)$ is computed and checked against each child's distance interval: $R_1$ is skipped (no intersection), the internal node is visited (intersection), and $R_{\text{res}}^{(\text{root})}$ is visited unconditionally. At the internal node, both $R_2$ and $R_3$ are visited. The right side of the figure expands $R_3$'s pivot table $P_{R_3}$, which holds distances from three historical queries ($p_1{=}q^8$, $p_2{=}q^{21}$, $p_3{=}q^{35}$). For each active pivot $p$, RECAST computes $d(q, p)$ once and prunes every object $x$ satisfying $|d(x, p) - d(q, p)| > r$. A dash in the table indicates a missing distance; such objects pass the pivot unchecked and require exact distance verification.

\hi{Query flow.}
A query $(q, r)$ enters at the root region. If the region has been partitioned with split center $s$ and child distance intervals $I_{\text{left}}$ and $I_{\text{right}}$, RECAST computes $d(q, s)$ and skips a child whose interval does not intersect $[d(q,s)-r, d(q,s)+r]$. The residual child has no interval, so it is always visited. If the region has not been partitioned, RECAST processes it as a leaf (Algorithm~\ref{alg:processleaf}). The query may then trigger a partition of the region (Section~\ref{sec:method:dispatch}). Algorithm~\ref{alg:query} summarizes the routing.

\begin{algorithm}[t]
\caption{Query routing}\label{alg:query}
\begin{algorithmic}[1]
\Procedure{Query}{$R, q, r, \mathit{result}$}
  \If{$R$ is partitioned with split center $s$}
    \State compute $d(q, s)$
    \ForAll{child $C \in \{R_{\text{left}}, R_{\text{right}}\}$}
      \If{$I_C \cap [d(q,s)-r,\; d(q,s)+r] \neq \emptyset$}
        \State \Call{Query}{$C, q, r, \mathit{result}$}
      \EndIf
    \EndFor
    \State \Call{Query}{$R_{\text{residual}}, q, r, \mathit{result}$}
  \Else
    \State \Call{ProcessLeaf}{$R, q, r, \mathit{result}$}
  \EndIf
\EndProcedure
\end{algorithmic}
\end{algorithm}

\hi{Recursive decomposition.}
A partition produces three children: left and right hold objects with known distances to the split center; the residual child holds objects whose distance to the split center was not computed. The residual child is a full region with its own table and can be further partitioned. Section~\ref{sec:method:dispatch} describes the partition mechanism.

%----------------------------------------------------------
\subsection{Signal-Driven Paid Pivot Management}\label{sec:method:region}
%----------------------------------------------------------

Observation~1 tells us that distances computed during a query can serve as reusable paid distances. The question is how to manage these distances within a region. Keeping all distances forever fills the table with old or low-value pivots, while a fixed eviction policy such as FIFO or LRU cannot fit both stable workloads, where good pivots should be preserved, and shifting workloads, where stale pivots should be replaced quickly. We use the same region cost signal to govern pivot admission, retention, and eviction, so that pivot management and the partition decision in Section~\ref{sec:method:dispatch} share a single signal-driven framework.

\hi{Object-level pruning.}
When a region $R$ is processed as a leaf, RECAST prunes candidates using $P_R$. Each pivot $p \in P_R$ is a past query center or a split center. The table stores $d(x, p)$ for each object $x$ whose distance to $p$ was computed by an earlier query. For a new query $(q, r)$, RECAST computes $d(q, p)$ once and prunes every object $x$ for which $|d(x, p) - d(q, p)| > r$. An object survives only if it passes every active pivot. Because $P_R$ is scoped to $R$, the pivots come from queries that visited $R$, not from queries elsewhere.

\hi{Admission and confirmation.}
Each region $R$ maintains a paid pivot table with a total budget of $K$ pivots. A new distance column enters the table only when the current query checked at least $c_{\min}$ objects and produced at least $f_{\min}$ false positives in $R$. The column starts in a \textit{trial} state. Each subsequent query that visits $R$ activates the column and records its pruning gain. Let $g_p$ denote the cumulative number of objects pruned by pivot $p$ minus the cumulative number of pivot-query distances $d(q, p)$ it consumed. When $g_p$ exceeds a confirmation threshold $\tau$, the pivot transitions to a \textit{confirmed} state.

\hi{Signal-driven eviction.}
When the table budget $K$ is exceeded or the number of trial pivots exceeds the trial cap $K_n$, the eviction policy is selected by the current cost signal in $R$:
\begin{itemize}
  \item When $c_R(q) < \bar{c}_R$ (the cost is falling and $R$ is still learning), or when $R$ is visited for the first time and no prior cost history exists, only trial pivots with $g_p \leq 0$ are evicted; confirmed pivots are kept.
  \item When $c_R(q) > \kappa \cdot \bar{c}_R$ ($\kappa{=}4$, i.e., a fourfold cost spike), pivots are evicted in order of oldest birth query, and confirmed pivots may also be evicted to make room for new distances.
  \item Otherwise, the pivot with the lowest eviction score $e_p = g_p / |V_p|$ is removed, where $|V_p|$ is the number of objects with a stored distance to $p$; trial pivots receive an additional penalty so that unproven columns are removed before confirmed ones.
\end{itemize}
The conditions that select the eviction behavior are the same cost observations used by the partition trigger in Section~\ref{sec:method:dispatch}: stable workloads keep good pivots and avoid partition, shifting workloads replace stale pivots and trigger partition candidates, and intermediate states fall back to gain-ordered decisions in both subsystems.
We use $K{=}32$ pivots per region with a trial cap of $K_n{=}8$; Section~\ref{sec:exp:settings} lists the full parameter setting.

Algorithm~\ref{alg:processleaf} details the leaf processing. It selects up to $K$ active pivots by gain (line~1), computes $d(q, p)$ for each (line~2), and prunes objects whose stored distance violates the triangle bound (lines~5--8). Survivors are verified by exact distance (lines~10--12). If the checked count and false-positive count exceed their thresholds, the distances from this query are recorded as a new trial column (lines~14--15). Finally, the cost signal selects the eviction behavior and the region statistics are updated (lines~17--19).

\begin{algorithm}[t]
\caption{Leaf processing}\label{alg:processleaf}
\begin{algorithmic}[1]
\Procedure{ProcessLeaf}{$R, q, r, \mathit{result}$}
  \State $\mathit{active} \gets$ select up to $K$ pivots from $P_R$ by gain $g_p$
  \ForAll{$p \in \mathit{active}$} compute $d(q, p)$ \EndFor \Comment{$K$ pivot-query distances}
  \ForAll{$x \in O_R$}
    \State $\mathit{pruned} \gets \mathit{false}$
    \ForAll{$p \in \mathit{active}$ with $d(x,p)$ stored}
      \If{$|d(x,p) - d(q,p)| > r$}
        \State $\mathit{pruned} \gets \mathit{true}$; \textbf{break}
      \EndIf
    \EndFor
    \If{not $\mathit{pruned}$}
      \State compute $d(q, x)$
      \If{$d(q, x) \leq r$} add $x$ to $\mathit{result}$ \EndIf
    \EndIf
  \EndFor
  \If{checked count $\geq c_{\min}$ and false positives $\geq f_{\min}$}
    \State record $\{d(q,x)\}$ as a new trial column in $P_R$
  \EndIf
  \State select eviction mode by $c_R(q)$ vs $\bar{c}_R$; confirm and evict accordingly
  \State update $H_R$: checked count, false positives, pivot pruned
  \State evaluate partition candidate (Section~\ref{sec:method:dispatch})
\EndProcedure
\end{algorithmic}
\end{algorithm}

%----------------------------------------------------------
\subsection{Cost-Signal-Driven Partition}\label{sec:method:dispatch}
%----------------------------------------------------------

Observations~1 and~3 together tell us that the distances computed during a query both produce reusable paid distances and measure structure fitness. Section~\ref{sec:method:region} described how the same signal drives pivot retention. This subsection describes how the same signal drives the partition timing. Partitioning a region introduces a routing distance on every future query and changes the structure that later queries depend on. A premature partition disrupts learning. A delayed partition keeps paying for a poor structure.

\hi{Split mechanism.}
A partition uses the triggering query $q$ as the split center $s$. Let $C_q \subseteq O_R$ be the objects whose distance to $s$ was computed during this query. RECAST computes the median $\mu$ of $\{d(x,s) : x \in C_q\}$ and assigns objects with $d(x,s) \leq \mu$ to the left child and objects with $d(x,s) > \mu$ to the right child. Let $L = \{x \in C_q : d(x,s) \leq \mu\}$ and $R' = \{x \in C_q : d(x,s) > \mu\}$. The distance intervals are $I_{\text{left}} = [\min_{x \in L} d(x,s),\; \mu]$ and $I_{\text{right}} = (\mu,\; \max_{x \in R'} d(x,s)]$. These intervals are fixed at split time and not updated afterward. Objects in $O_R \setminus C_q$, whose distance to $s$ was not computed, are collected into the residual child. The split center $s$ is installed as a pivot in the left and right children, providing them with an initial pruning column at no extra cost.

\hi{Query statistics.}
Each region $R$ maintains statistics $H_R$ that track the recent behavior of queries in $R$. These include the number of visits, the exponential moving average of the checked-object count $\bar{c}_R$, the total false positives, and the total objects pruned by the pivot table. The EMA is updated after each visit as $\bar{c}_R \leftarrow \lambda \cdot c_R(q) + (1-\lambda) \cdot \bar{c}_R$, where $\lambda$ controls how fast the average responds to new queries. These statistics are updated at the end of each leaf processing step (Algorithm~\ref{alg:processleaf}, line 19).

\hi{Candidate selection.}
A region $R$ becomes a partition candidate for the current query $(q, r)$ when the following conditions are all met:
\begin{enumerate}
  \item $|O_R| \geq \sigma$ ($\sigma{=}512$);
  \item $R$ has been visited at least twice;
  \item the checked-object count $c_R(q)$ satisfies $c_R(q) / |O_R| \geq \alpha$;
  \item the false-positive ratio $f_R(q) / c_R(q) \geq \beta$;
  \item at least one of the following holds:
    \begin{enumerate}
      \item[(5a)] the pivot prune rate $\mathit{pruned}_R(q) / (c_R(q) + \mathit{pruned}_R(q)) < \rho$ (the existing pivots are not pruning well, suggesting the structure is saturated for this region);
      \item[(5b)] the checked count exceeds $\kappa \cdot \bar{c}_R$ ($\kappa{=}4$; a fourfold cost spike suggests the workload has shifted to a new region);
      \item[(5c)] the pivot table is full (the region has no room for new distances, so further learning within the current scope is limited).
    \end{enumerate}
\end{enumerate}
A region whose cost is falling ($c_R(q) < \bar{c}_R$) is not a candidate, because the pivot table is still learning. The three trigger conditions (5a saturation, 5b spike, 5c table full) correspond directly to the three eviction modes in Section~\ref{sec:method:region}: 5a triggers when gain-based eviction in $R$ no longer recovers pruning power, 5b coincides with aggressive-mode eviction when the workload shifts, and 5c arises when conservative-mode eviction cannot free space, so all adaptive decisions in RECAST derive from the same cost observation.

\hi{Shadow validation.}
A candidate partition is not committed at once. RECAST creates a shadow split with center $s$ and median $\mu$, and tests it against subsequent queries without changing the actual structure. For each subsequent query $(q', r')$ that visits $R$, the shadow computes $d(q', s)$ and determines which child would have been skipped. It records the \textit{counterfactual saved checks}: the number of objects in the skippable child that were not already pruned by the existing pivot table. The net score is
\[
  \mathit{net} = \sum_{i} \mathit{saved}_i - \sum_{i} \mathit{routing}_i,
\]
where $\mathit{saved}_i$ is the counterfactual saved checks of the $i$-th subsequent query and $\mathit{routing}_i = 1$ is the routing distance it would have added. The partition is committed when the shadow has been tested for at least $v$ queries and $\mathit{net} \geq \eta$. If $\mathit{net} < \eta$ after $v$ queries, the shadow is discarded.

This two-stage process follows from Observation~3. Candidate selection reads the cost signal to detect that a region may need reorganization. Shadow validation verifies that the specific partition would reduce distance computations. Together they avoid both premature and delayed partition. The shadow test also mitigates the risk of a poor split center: if the triggering query happens to be an outlier, the shadow will not accumulate enough saved checks to pass the commit threshold.
In Figure~\ref{fig:architecture}, the partition at the internal node (split center $q^{12}$) was committed only after the shadow test confirmed that routing through $q^{12}$ would save enough distance computations in subsequent queries to offset the added routing cost.

Algorithm~\ref{alg:partition} summarizes the partition decision. It is invoked at the end of each leaf processing step (Algorithm~\ref{alg:processleaf}, line~20). When the candidate conditions are met, a shadow is created; when a shadow passes validation, the region is committed into three children.

\begin{algorithm}[t]
\caption{Partition decision}\label{alg:partition}
\begin{algorithmic}[1]
\Procedure{EvaluatePartition}{$R, q, r, C_q$}
  \If{$R$ has an active shadow $S$}
    \State compute $d(q, S.\mathit{center})$
    \State $\mathit{saved} \gets$ objects in skippable child not already pruned
    \State $S.\mathit{net} \gets S.\mathit{net} + \mathit{saved} - 1$
    \State $S.\mathit{visits} \gets S.\mathit{visits} + 1$
    \If{$S.\mathit{visits} \geq v$ and $S.\mathit{net} \geq \eta$}
      \State \Call{CommitPartition}{$R, S$} \Comment{create 3 children}
    \ElsIf{$S.\mathit{visits} \geq v$ and $S.\mathit{net} < \eta$}
      \State discard $S$
    \EndIf
  \ElsIf{candidate conditions (1)--(5) are met}
    \State $\mu \gets \mathrm{median}\{d(x, q) : x \in C_q\}$
    \State create shadow $S$ with $S.\mathit{center} \gets q$,\; $S.\mu \gets \mu$,\; $S.C \gets C_q$,\; $S.D \gets \{d(x,q): x \in C_q\}$
  \EndIf
\EndProcedure
\Statex
\Procedure{CommitPartition}{$R, S$}
  \State $L \gets \{x \in S.C : S.D[x] \leq S.\mu\}$
  \State $R' \gets \{x \in S.C : S.D[x] > S.\mu\}$
  \State $U \gets O_R \setminus S.C$ \Comment{residual: distances unknown}
  \State create $R_{\text{left}}$ with objects $L$, install $S.\mathit{center}$ as pivot
  \State create $R_{\text{right}}$ with objects $R'$, install $S.\mathit{center}$ as pivot
  \State create $R_{\text{residual}}$ with objects $U$
  \State set $I_{\text{left}} \gets [\min_{x \in L} S.D[x],\; S.\mu]$,\; $I_{\text{right}} \gets (S.\mu,\; \max_{x \in R'} S.D[x]]$
  \State mark $R$ as partitioned
\EndProcedure
\end{algorithmic}
\end{algorithm}

\hi{Cost and correctness.}
Each stored distance $d(x, p)$ is recorded once. With a budget of $K$ pivots per region, the worst-case storage is $O(n \cdot K)$ distance pairs; in practice, only objects that have been checked by at least one query have stored distances. Each query pays one routing distance per visited internal region, up to $K$ pivot-query distances per visited leaf, and exact distances for survivors. Regions smaller than $\sigma$ are not further partitioned, which limits the tree depth in practice. RECAST prunes an object only when a stored distance and the triangle inequality prove it cannot be in the result. Objects whose distance to a pivot is missing pass the pivot unchecked. The residual child is always visited because it has no distance interval. Therefore, every query returns the exact answer.

%==========================================================
\section{Experiments}\label{sec:exp}
%==========================================================

We evaluate RECAST against adaptive and pre-built baselines.
We first compare RECAST with AV-tree across multiple workload patterns (Section~\ref{sec:exp:main}), then include pre-built indexes in a cumulative-cost analysis that accounts for their upfront build cost (Section~\ref{sec:exp:cumulative}).
Section~\ref{sec:exp:ablation} studies the contribution of each component.

%----------------------------------------------------------
\subsection{Experimental Settings}\label{sec:exp:settings}
%----------------------------------------------------------

\hi{Datasets.}
Table~\ref{tab:datasets} summarizes the five datasets.

\begin{table}[t]
\centering
\caption{Datasets used in the experiments.}
\label{tab:datasets}
\small
\begin{tabular}{lrrll}
\toprule
Dataset & Size & Dim & Distance & Note \\
\midrule
glove100   & 1.18M & 100 & $L_2$ & word embeddings \\
sift1m     & 1.0M  & 128 & $L_2$ & SIFT descriptors \\
sift10m    & 10.0M & 128 & $L_2$ & scalability \\
colors112d & 0.11M & 112 & QFD   & non-$L_2$ metric \\
nasa20d    & 0.4M  &  20 & $L_2$ & low-dim boundary \\
\bottomrule
\end{tabular}
\end{table}

\hi{Workloads.}
We use four workload patterns, each consisting of 1\,000 queries at 1\% selectivity:
a \emph{fixed} pattern where all queries target the same region,
a \emph{jump} where the focus shifts from region A to region B at query 500,
a \emph{drift} that gradually moves from A to B over the full trace,
and a \emph{three-jump} that cycles A$\to$B$\to$C.

\hi{Baselines.}
\emph{Adaptive:} AV-tree~\cite{lampropoulos2023avtree} (cracking threshold $\theta{=}128$, distance caching enabled).
\emph{Pre-built:} LAESA (32 pivots)~\cite{mico1994laesa} and GNAT~\cite{brin1995gnat}.
\emph{Reference:} Linear scan.
We fix $K{=}32$ for RECAST throughout (both pivot budget and active budget).

\hi{Metrics.}
The number of distance computations per query is the primary measure; query time is the secondary measure.
Each experiment runs three seeds; we report the mean.

\hi{Hardware and implementation.}
All methods are implemented in C++ and compiled with g++ 11.4.0 using \texttt{-O3 -mavx -march=native}. Experiments run on a 754\,GiB Ubuntu 22.04.5 LTS machine with two Intel Xeon Platinum 8276L CPUs at 2.20\,GHz.

%----------------------------------------------------------
\subsection{Main Comparison}\label{sec:exp:main}
%----------------------------------------------------------

This experiment tests whether RECAST reduces per-query distance computations consistently across datasets and workload patterns. We compare RECAST with AV-tree across all five datasets and four workloads (20 combinations).
Both are adaptive indexes that build from queries, so per-query distance computations (DC) and query time are directly comparable without a build-cost offset.
Linear scan is included as a sanity reference.

\begin{figure*}[t]
\centering
\includegraphics[width=\textwidth]{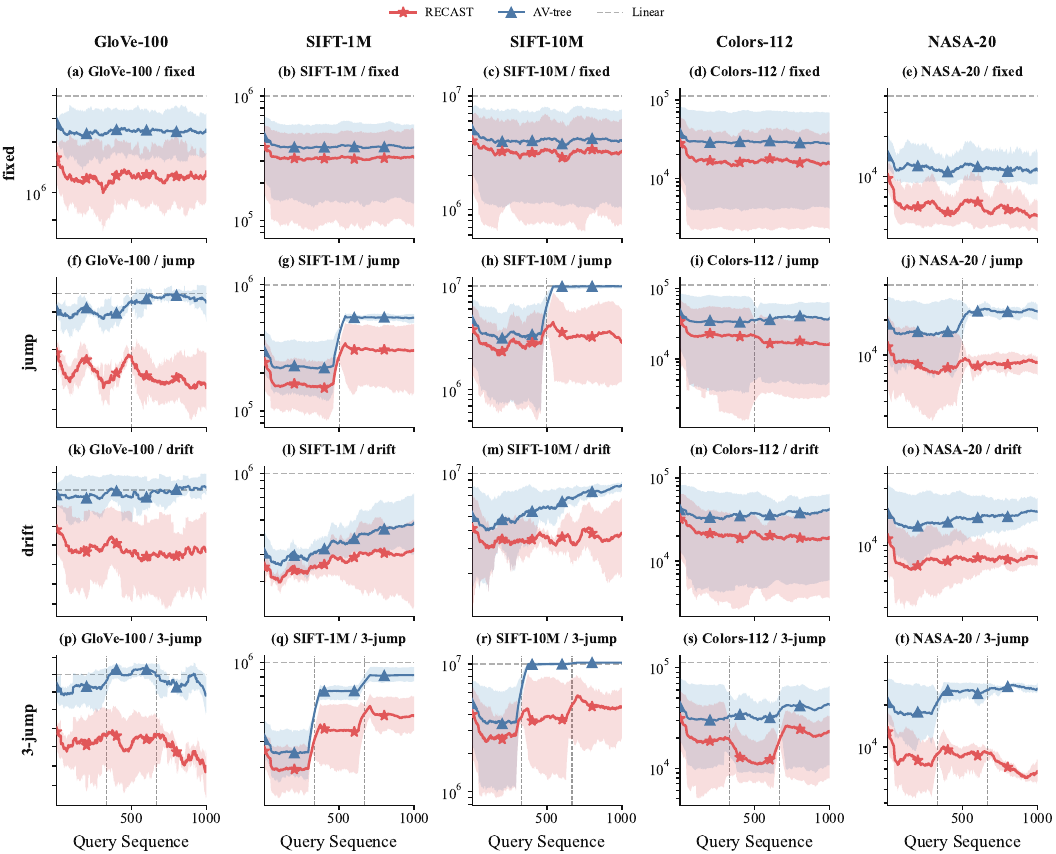}
\caption{Per-query distance computations on five datasets under four workloads.}
\label{fig:per_query}
\end{figure*}

Overall, RECAST reduces per-query distance computations to $0.615\times$ that of AV-tree (geometric mean across all 20 combinations) and query time to $0.863\times$, winning all 20 combinations in distance computations.
RECAST also reduces query time on 16 of 20 combinations; on nasa20d ($1.7$--$2.3\times$ slower), the per-pivot management overhead exceeds the distance-computation savings because individual distance evaluations in 20 dimensions cost only ${\sim}0.5\,\mu$s.
Figure~\ref{fig:per_query} plots the per-query distance computations across all 20 combinations.

\hi{Cross-dataset comparison.}
The DC advantage varies across datasets in a pattern that reflects dimensionality, dataset size, and the cost of the distance function.
On nasa20d (20 dimensions), RECAST achieves the largest DC reduction (geometric mean $0.435\times$ across four workloads) (Figure~\ref{fig:per_query}, panels e, j, o, t), because the low intrinsic dimensionality makes each stored pivot distance a strong discriminator: the triangle-inequality lower bound is tight relative to the search radius, so a small number of region-local pivots already prunes most candidates.
On sift10m ($n{=}10$M, 128 dimensions), the per-dataset geometric mean is $0.569\times$, and on colors112d (112 dimensions, quadratic-form distance) it is $0.558\times$.
The sift10m result confirms that the DC advantage scales with dataset size: at $10\times$ the objects of sift1m, the per-query saving from accumulated pivot distances grows because more candidates fall within each region and each pivot column amortizes over a larger object set.
The colors112d result confirms that the region-scoped design is metric-agnostic; the quadratic-form distance is not $L_2$, yet the same pivot-reuse mechanism reduces DC by a comparable factor.
On glove100 (100-dimensional word embeddings), the DC ratio is $0.926\times$, the smallest improvement among all datasets (panels a, f, k, p). We attribute this to the distance concentration in the embedding space: the pairwise distance distribution of glove100 is narrower than those of sift1m and colors112d, which makes the triangle-inequality lower bound loose for most object-pivot pairs regardless of the pivot's position. RECAST still reduces DC on glove100 because region-scoping restricts each pivot table to a local neighborhood where the effective distance spread is wider than the global distribution, but the absolute gain is smaller.
On sift1m (128 dimensions, $n{=}1$M), the geometric mean is $0.685\times$, intermediate between the low-dimensional and the highly concentrated cases.

\hi{Workload comparison.}
Under the fixed workload, both adaptive methods converge because neither faces a shift.
The gap widens under jump and drift workloads: after the query focus moves to a new region, AV-tree's existing tree structure provides little pruning in the new region, whereas RECAST's region-local tables in the new target may already hold partial distances from earlier visits.
Quantitatively, the per-workload DC geometric means (across five datasets) are: fixed $0.707\times$, drift $0.639\times$, jump $0.572\times$, and three-jump $0.552\times$.
The three-jump workload produces the largest advantage (panels p--t) because each shift exposes AV-tree to a region where its existing tree structure provides little pruning, and the single cached distance per object limits how quickly it can re-learn.
RECAST's region-local tables, by contrast, retain distances accumulated during phase A that partially transfers when the workload returns to a nearby region or when the residual child from an earlier split already covers objects in the new target.
Under fixed, both methods converge to a stable structure and the gap is smallest; the remaining $0.707\times$ reduction comes from RECAST's multi-pivot pruning within each region, which AV-tree cannot match with its single cached distance per object.

Figure~\ref{fig:per_query} visualizes the per-query behavior across all 20 combinations.
Under jump and three-jump workloads, AV-tree's per-query distance computations spike at the shift point and recover slowly; RECAST's spike is shallower and recovery is faster, because region-local pivot tables retain distances from earlier visits.
Under drift, the transition is gradual and RECAST maintains a consistently lower envelope.
On nasa20d, the distance-computation gap is large (DC ratio $0.435\times$), yet RECAST's query time exceeds AV-tree's by $1.7$--$2.3\times$ because each distance evaluation in 20 dimensions costs only ${\sim}0.5\,\mu$s, and the per-pivot overhead outweighs the saved computations.
This boundary case confirms that RECAST targets settings where distance computations are expensive relative to the bookkeeping cost.

%----------------------------------------------------------
\subsection{Cumulative Cost versus Pre-Built Indexes}\label{sec:exp:cumulative}
%----------------------------------------------------------

This experiment tests whether RECAST's cumulative cost remains competitive with pre-built indexes that pay their build cost upfront. Pre-built indexes pay a one-time build cost and then answer every query at a fixed per-query rate.
We include this build cost in the cumulative DC curve so that the trade-off between upfront investment and adaptive learning is visible.
We add the construction cost of each pre-built index to its cumulative curve before the first query.

\begin{figure*}[t]
\centering
\includegraphics[width=\textwidth]{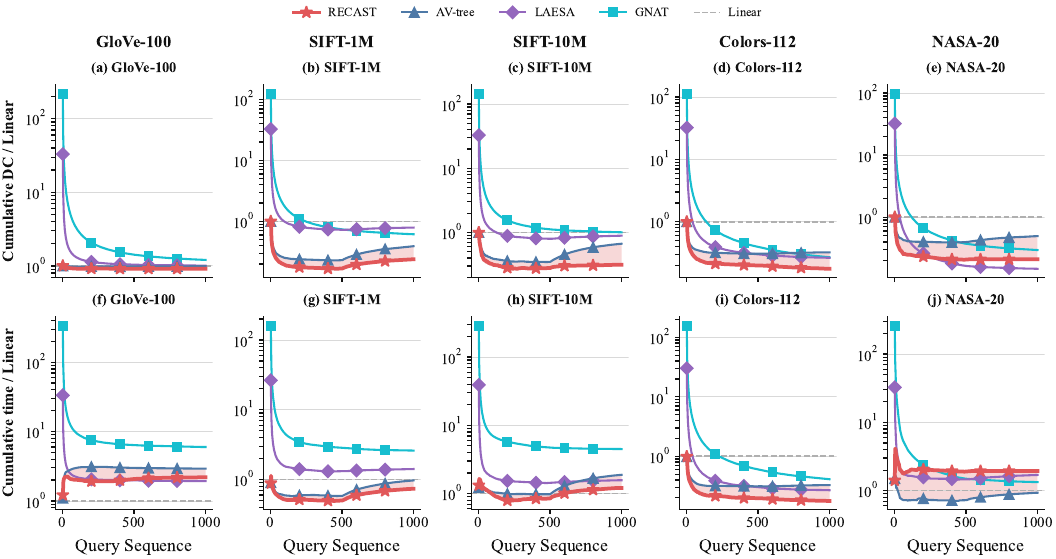}
\caption{Cumulative cost normalized to linear scan (jump workload). Top: distance computations. Bottom: query time. Pre-built methods include build cost.}
\label{fig:cumulative}
\end{figure*}

Figure~\ref{fig:cumulative} plots the cumulative cost normalized to linear scan.

\hi{Break-even analysis.}
Pre-built indexes pay their full construction cost before answering any query, so their cumulative curves start above zero.
Adaptive indexes start from zero but pay a higher per-query cost during early learning.
The break-even point is the query index at which the adaptive curve crosses below the pre-built curve.
On sift10m (Figure~\ref{fig:cumulative}(c)), LAESA's build cost is $320\times 10^6$ DC ($32 \times n$).
RECAST's cumulative DC curve starts from zero and remains below LAESA's curve for the entire 1\,000-query trace, because the build cost alone exceeds the total distance computations RECAST pays over 1\,000 queries.
On sift1m (build cost $32\times 10^6$ DC), RECAST's cumulative curve crosses below LAESA within the first 150 queries; once region-local pivots accumulate, the per-query DC drops and the gap widens steadily.
On glove100 (build cost $38.4\times 10^6$ DC), the crossover also occurs early because RECAST's per-query DC (${\sim}1.09\times 10^6$) is only marginally below LAESA's (${\sim}1.17\times 10^6$), but the build cost deficit that LAESA carries is never recovered within 1\,000 queries.
On nasa20d (Figure~\ref{fig:cumulative}(e), build cost $1.28\times 10^7$ DC), LAESA's 32 global pivots already prune well in 20 dimensions (per-query DC ${\sim}4.6\times 10^3$), lower than both AV-tree ($19.9\times 10^3$) and RECAST ($8.5\times 10^3$).
LAESA's cumulative curve crosses below RECAST's after approximately 200 queries and remains lower for the rest of the trace.
This is the one dataset where a pre-built index outperforms RECAST in cumulative DC, because the low dimensionality makes a small set of global pivots sufficient and the build cost is modest relative to $n$.
On colors112d (build cost $3.6\times 10^6$ DC), RECAST crosses below LAESA at approximately query 200.
Across all datasets, AV-tree's cumulative curve lies between RECAST and LAESA, confirming that region-scoped distance reuse, not merely adaptive construction, drives the cumulative improvement over both pre-built and adaptive baselines.
We show the jump workload because it is the standard single-shift pattern; trends under fixed, drift, and three-jump are consistent.

%----------------------------------------------------------
\subsection{Component Ablation}\label{sec:exp:ablation}
%----------------------------------------------------------

\begin{figure}[t]
\centering
\includegraphics[width=\columnwidth]{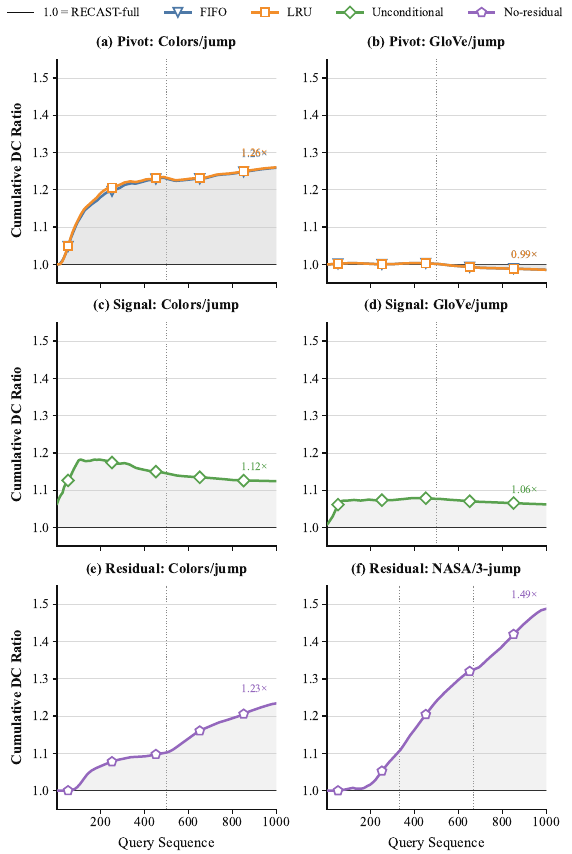}
\caption{Component ablation. Each panel compares RECAST-full (red) with an ablated variant on cumulative distance computations.}
\label{fig:ablation}
\end{figure}

This experiment isolates the contribution of each RECAST component by replacing it with a simpler alternative while keeping the rest intact. Figure~\ref{fig:ablation} plots the cumulative distance-computation ratio (variant / RECAST-full) over the query trace: a line above 1.0 means removing the component increases cost.

\hi{Pivot management.}
We replace the signal-driven adaptive eviction policy (Section~\ref{sec:method:region}) with two fixed alternatives: FIFO (evict the oldest pivot) and LRU (evict the least recently used pivot), keeping the same pivot budget $K{=}32$.
On colors112d (Figure~\ref{fig:ablation}(a)), FIFO and LRU increase cumulative distance computations by 20--30\%. The QFD distance function concentrates pruning power in a few pivots whose triangle-inequality bounds align well with the query distribution; the adaptive policy detects these pivots through their high cumulative gain $g_p$ and retains them, while FIFO evicts them once they become the oldest entry regardless of pruning value.
On glove100 (Figure~\ref{fig:ablation}(b)), the gap is small ($<$5\%) because distance concentration limits all pivots to similar marginal pruning power.

\hi{Signal and shadow validation.}
We replace the cost-signal-driven partition trigger with shadow validation (Section~\ref{sec:method:dispatch}) by an unconditional split: every qualifying query triggers an immediate split without cost-signal gating or shadow testing.
On colors112d (Figure~\ref{fig:ablation}(c)), unconditional splitting increases cumulative distance computations by 10--15\%, because it fractures regions whose pivot tables are still learning, destroying accumulated distances before it reaches full pruning effectiveness.
On glove100 (Figure~\ref{fig:ablation}(d)), the distance-computation increase is smaller (5--10\%), but query time roughly doubles because each unnecessary split adds one routing distance computation $d(q, s)$ on every subsequent query that traverses the over-fragmented tree.

\hi{Residual child.}
We remove the residual child by assigning all unclassified objects to the parent bucket instead of creating an independent child region.
This produces the largest ablation effect. On nasa20d under three-jump (Figure~\ref{fig:ablation}(f)), the cumulative ratio reaches $1.5\times$ by query 1\,000. Each of the three workload shifts routes queries into a region where unclassified objects from an earlier split reside. With the residual child, these objects form an independent region and begin accumulating pivots from the new query stream. Without it, they remain in the parent bucket, which mixes objects from all prior splits; every visit scans them linearly because the parent's pivot table was built for the old workload and provides little pruning for the new one.

\begin{figure}[t]
\centering
\includegraphics[width=\columnwidth]{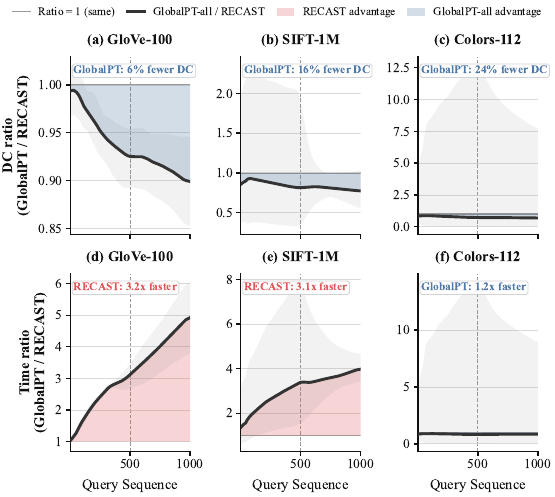}
\caption{Region scoping vs.\ unbounded global table (jump workload). Top: cumulative distance computations. Bottom: cumulative query time.}
\label{fig:region_scoping}
\end{figure}

\hi{Region scoping.}
We compare RECAST with GlobalPT-all, a single unbounded global pivot table that retains every query column with no region scoping and no budget limit.
Figure~\ref{fig:region_scoping} overlays cumulative distance computations (solid lines) and cumulative query time (dashed lines) for both methods.
On glove100 and sift1m, GlobalPT-all's solid line (distance computations) is slightly below RECAST's, but its dashed line (query time) is far above.
RECAST accepts a bounded increase in distance computations (geometric mean $1.29\times$) in exchange for substantially lower query time (geometric mean $0.39\times$) because scanning a per-region table of $K{=}32$ columns is far cheaper than scanning a global table that grows to 1\,000 columns.
On colors112d ($n{=}112$K), the dataset is small enough that GlobalPT-all's table scan does not dominate, and query times are comparable.

%==========================================================
\section{Related Work}\label{sec:rw}
%==========================================================

We organize related work by whether the index uses query information during construction: methods that build entirely from the data (Section~\ref{sec:rw:data}), and methods that build as a side effect of answering queries (Section~\ref{sec:rw:adaptive}).

\subsection{Data-Driven Indexes for Exact Similarity Search}\label{sec:rw:data}

Classical multidimensional indexes such as the KD-tree~\cite{bentley1975kdtree} and the R-tree~\cite{guttman1984rtree} partition a coordinate space with axis-aligned or rectangular regions, and iDistance~\cite{jagadish2005idistance} maps points to a one-dimensional B$^+$-tree key for nearest-neighbor search. These methods assume that objects carry coordinates. In a general metric space, objects may have no coordinates and an index can prune only through the triangle inequality, using precomputed distances to a set of reference objects~\cite{chavez2001searching,chen2022survey}.

Pivot- and partition-based metric space indexes follow this principle. AESA~\cite{vidal1986aesa} stores all pairwise distances and achieves near-constant query cost on small datasets; LAESA~\cite{mico1994laesa} reduces the storage to a fixed set of pivots, and later work studies how to select pivots that maximize pruning power~\cite{chen2017pivot}, often through facility-location heuristics akin to the $k$-center problem~\cite{hochbaum1985kcenter}. Tree-structured indexes partition the space around reference objects: VP-tree~\cite{yianilos1993vptree} and MVP-tree~\cite{bozkaya1999mvptree} split by vantage points, GNAT~\cite{brin1995gnat} uses multiple split points per node, SAT~\cite{navarro2002sat} navigates by spatial approximation, the M-tree~\cite{ciaccia1997mtree} and PM-tree~\cite{skopal2004pmtree} support balanced disk-based access, and cover trees~\cite{beygelzimer2006cover} and rank cover trees~\cite{houle2013rankcover} give worst-case guarantees under bounded intrinsic dimension. Clustering-based decompositions such as the list of clusters~\cite{chavez2005loc} and the scalable M-index~\cite{novak2011mindex} group nearby objects to bound the search to a few clusters. Learned indexes such as LIMS~\cite{tian2022lims} use data clustering and pivot-based feature transformations to reduce the candidate set for exact search. Specialized indexes target exact similarity search over data series: SAX-based summaries~\cite{camerra2010isax} and variable-length indexes~\cite{linardi2018ulisse} yield tight lower bounds for pruning, recent systems push exact search to scale~\cite{echihabi2022hercules}, and progressive variants return refining answers with quality guarantees~\cite{gogolou2023pros}. In all of these, the partition is fixed at build time from the data distribution; a region that is queried heavily and a region that is never touched receive the same indexing effort, and the structure cannot redirect that effort toward the part of the space the workload actually visits.

Some metric space indexes support data updates. EGNAT~\cite{uribe2006egnat}, dynamic spatial approximation trees~\cite{navarro2008dsat}, and the dynamic VP-tree~\cite{fu2000dynvptree} allow incremental insertions and deletions, and D-Cache~\cite{skopal2012dcache} reuses previously computed distances across queries for pruning. These mechanisms accommodate changes in the \emph{data}, but the partition structure still reflects the data distribution rather than the query distribution, and is not reorganized when queries concentrate in a particular region.

A separate line of work trades exactness for speed: when distances concentrate in high dimensions and exact pruning weakens~\cite{beyer1999nn}, approximate nearest-neighbor methods such as locality-sensitive hashing~\cite{gionis1999lsh} and proximity graphs~\cite{malkov2020hnsw} accept a small recall loss for sublinear query time. RECAST targets exact range search, where every object within the query radius must be returned, so these approximate methods are complementary rather than competing.

All of these methods build a fixed structure before any query arrives. In our setting, the index starts empty and must learn its structure from the queries themselves; the query distribution may also shift during execution, which a fixed structure cannot track.

\subsection{Adaptive Indexing}\label{sec:rw:adaptive}

A second line of work builds the index incrementally as queries arrive, with no separate construction phase.

The idea of building structure lazily in response to queries dates back to deferred data structuring~\cite{karp1988deferred}. Database cracking~\cite{idreos2007cracking} brought it to relational columns: each query partitions an unsorted array into progressively finer sorted segments, in effect an incremental quicksort~\cite{hoare1961quicksort} driven by the query predicates, and self-organizing reconstruction extends it to full tuples~\cite{idreos2009selforg}. A large body of work refined the relational case, adding support for updates~\cite{idreos2007updating}, convergence through merging~\cite{idreos2011merging,holanda2021mergesort}, robustness to adversarial query orders~\cite{halim2012stochastic}, efficient scan-based reorganization~\cite{pirk2014cracking}, holistic integration with the query optimizer~\cite{petraki2015holistic}, self-tuning index selection~\cite{graefe2010selftuning}, variants over encrypted data~\cite{karras2016encrypted}, and systematic analyses of the overhead each query pays~\cite{schuhknecht2013uncracked,schuhknecht2016analysis}. A complementary line bounds the convergence of cracking theoretically~\cite{zardbani2020revisiting} and smooths its early-query cost through progressive indexing that spreads the sorting effort across a controlled number of queries~\cite{holanda2019progressive,jensen2021revisiting}. These refinements all target the relational, coordinate-ordered setting, where a total order over key values makes the partition boundaries explicit.

Later work extended adaptive indexing to multidimensional coordinate data: cracking KD-trees~\cite{holanda2018crackingkd}, the query-aware spatial index QUASII~\cite{pavlovic2018quasii}, multidimensional adaptive and progressive indexes~\cite{nerone2021multidim}, adaptive indexes over objects with spatial extent~\cite{zardbani2023spatial}, and recent work on updating such indexes~\cite{zardbani2025glide}. These methods rely on hyperplane- or coordinate-based partitioning that does not generalize to arbitrary metric spaces, where objects carry no coordinates and only distances are available. The same query-driven philosophy has reached similarity search in other domains: the adaptive data series index builds and refines its structure in response to incoming queries rather than in a separate construction phase~\cite{zoumpatianos2016ads}.

A parallel line of learned indexes replaces index nodes with models that predict key positions~\cite{kraska2018learned,galakatos2019fiting,ferragina2020pgm,kipf2020radixspline}, with later work supporting updates~\cite{ding2020alex}, multi-dimensional~\cite{nathan2020flood} and workload-skewed data~\cite{ding2020tsunami,marcus2020benchmark}, as part of a broader vision of self-driving database systems~\cite{pavlo2017selfdriving,kraska2019sagedb}. Like cracking, these methods assume a total order or coordinate representation and do not address the metric-space setting.

AV-tree~\cite{lampropoulos2023avtree} advances adaptive indexing from coordinate spaces to generic high-dimensional metric spaces, supporting both exact range and $k$-NN queries. It uses each query center as a vantage point, cracks the data array around the query ball, and caches each object's distance to its parent split pivot for triangle-inequality pruning in later queries. AV-tree demonstrates that the cracking principle can work in metric spaces and achieves per-query cost that converges toward that of pre-built indexes such as MVP-tree after enough queries.

AV-tree's design keeps at most one cached distance per object per leaf, so leaf-level pruning is bounded by a single pivot. Distances computed by earlier queries on ancestor pivots are not carried forward when a leaf is reorganized. This means that a costly query in a region produces distances that are used once for cracking but not accumulated as multi-pivot pruning distances for later queries in the same region. The index also treats every query uniformly: a query that produces reusable distances, a query that indicates the region is already well served, and a query that signals a workload shift all trigger the same cracking operation.

CrackIVF~\cite{mageirakos2025crackivf} applies cracking to approximate nearest-neighbor search over IVF indexes, demonstrating that adaptive construction scales to large vector datasets. CrackIVF focuses on the approximate search setting and explicitly scopes out query distribution shifts.

RECAST belongs to the adaptive group and differs from prior work in three ways: it partitions only when a region's cost signal and a shadow test agree the change pays off (Section~\ref{sec:method:dispatch}), rather than before any query~\cite{chen2017pivot} or on every qualifying query~\cite{lampropoulos2023avtree}; it accumulates a region-scoped multi-pivot table from already-paid distances without requiring coordinates (Section~\ref{sec:method:region}); and its residual child keeps every object in a learnable region (Section~\ref{sec:method:overview}).

%==========================================================
\section{Conclusion}\label{sec:conclusion}
%==========================================================

We have presented RECAST, a region-scoped adaptive index for exact similarity search that treats each query region as a recursive online task. By organizing the distances that queries are forced to compute into region-scoped pruning structures and reading their cost trends as adaptive signals, RECAST adapts its structure to the query stream without a separate construction phase. Experiments on five datasets under four workload patterns showed that RECAST reduced per-query distance computations on all 20 dataset-workload combinations and query time on 16 of 20 compared with AV-tree, with the largest gains under multi-shift workloads and diminishing returns when distance concentration limits pivot-based pruning or when individual distance evaluations are cheap. Ablation studies confirmed that each component contributes to the overall improvement. Extending the cost-signal framework to $k$NN queries, where the effective search radius shrinks dynamically, is a natural next step.

\section*{Acknowledgments}
This work is partially supported by the NSFC project 62532007.

\newpage

%==========================================================
% References
%==========================================================

\end{document}